\let\origvec\vec
\documentclass[runningheads]{llncs}
\newcommand{\korona}{\textsc{Korona}}
\newcommand{\KOI}{$\cal{KOI}$}
\newcommand{\KOIB}{$\cal{KOI}\;$}
\newcommand{\koronaB}{\textsc{Korona$\;$}}

\let\vec\origvec
\usepackage[T1]{fontenc}
\usepackage[utf8]{inputenc}
\usepackage{lmodern}
\usepackage{makeidx}  
\usepackage{graphicx}
\usepackage[caption=false]{subfig}
\usepackage{amsmath}
\newtheorem{mydef}{Definition}

\usepackage{array}
\newcolumntype{P}[1]{>{\centering\arraybackslash}p{#1}}
\usepackage{paralist}
\usepackage{multirow}
\usepackage{rotating}
\usepackage{floatrow}
\usepackage{colortbl}
\DeclareMathOperator*{\argmax}{argmax}
\usepackage{pgfplots}
\usepackage{pgfplotstable}
\usepackage{booktabs}
\usepackage{caption}
\usepackage{hyphenat}
\usepackage{hyperref}

\usepackage{listings}
\usepackage{comment}
\usepackage{xcolor,colortbl}
\usepackage{adjustbox}
\newenvironment{nospaceflalign*}
 {\setlength{\abovedisplayskip}{-3.5pt}\setlength{\belowdisplayskip}{4pt}%
  \csname flalign*\endcsname}
 {\csname endflalign*\endcsname\ignorespacesafterend}

\pgfplotstableset{
    /color cells/min/.initial=0,
    /color cells/max/.initial=1000,
    /color cells/textcolor/.initial=,
    color cells/.code={%
        \pgfqkeys{/color cells}{#1}%
        \pgfkeysalso{%
            postproc cell content/.code={%
                \begingroup
                \pgfkeysgetvalue{/pgfplots/table/@preprocessed cell content}\value
                \ifx\value\empty
                    \endgroup
                \else
                \pgfmathfloatparsenumber{\value}%
                \pgfmathfloattofixed{\pgfmathresult}%
                \let\value=\pgfmathresult
                \pgfplotscolormapaccess
                    [\pgfkeysvalueof{/color cells/min}:\pgfkeysvalueof{/color cells/max}]
                    {\value}
                    {\pgfkeysvalueof{/pgfplots/colormap name}}%
                \pgfkeysgetvalue{/pgfplots/table/@cell content}\typesetvalue
                \pgfkeysgetvalue{/color cells/textcolor}\textcolorvalue
                \toks0=\expandafter{\typesetvalue}%
                \xdef\temp{%
                    \noexpand\pgfkeysalso{%
                        @cell content={%
                            \noexpand\cellcolor[rgb]{\pgfmathresult}%
                            \noexpand\definecolor{mapped color}{rgb}{\pgfmathresult}%
                            \ifx\textcolorvalue\empty
                            \else
                                \noexpand\color{\textcolorvalue}%
                            \fi
                            \the\toks0 %
                        }%
                    }%
                }%
                \endgroup
                \temp
                \fi
            }%
        }%
    }
}

\newfloatcommand{capbtabbox}{table}[][\FBwidth]

\begin{document}
\title{Unveiling Scholarly Communities over Knowledge Graphs}
\author{Sahar Vahdati\inst{1}\orcidID{0000-0002-7171-169X} \and Guillermo Palma\inst{2}\orcidID{0000-0002-8111-2439} \and Rahul Jyoti Nath\inst{1} \and Christoph Lange\inst{1,4}\orcidID{0000-0001-9879-3827} \and S{\"o}ren Auer\inst{2,3}\orcidID{0000-0002-0698-2864} \and Maria-Esther Vidal\inst{2,3}\orcidID{0000-0003-1160-8727}}
\authorrunning{S. Vahdati et al.}
\institute{
University of Bonn, Germany
\email{\{vahdati,langec\}@cs.uni-bonn.de,s6ranath@uni-bonn.de,}\\
\and
L3S Research Center, Germany
\email{\{palma,auer,vidal\}@L3S.de} \\
\and
TIB Leibniz Information Centre for Science and Technology, Hannover, Germany
\email{Maria.Vidal@tib.eu}
\and
Fraunhofer IAIS, Germany
}
\maketitle 
\vspace*{-.75cm}
\begin{abstract}      
Knowledge graphs represent the meaning of properties of real-world entities and relationships among them in a natural way. 
Exploiting semantics encoded in knowledge graphs enables the implementation of knowledge-driven tasks such as semantic retrieval, query processing, and question answering, as well as solutions to knowledge discovery tasks including pattern discovery and link prediction. 
In this paper, we tackle the problem of knowledge discovery in scholarly knowledge graphs, i.e., graphs that integrate scholarly data, and present \korona, a knowledge-driven framework able to unveil scholarly communities for the prediction of scholarly networks. \koronaB implements a graph partition approach and relies on semantic similarity measures to determine relatedness between scholarly entities. As a proof of concept, we built a scholarly knowledge graph with data from researchers, conferences, and papers of the Semantic Web area, and apply \koronaB to uncover co-authorship networks. Results observed from our empirical evaluation suggest that exploiting semantics in scholarly knowledge graphs enables the identification of previously unknown relations between researchers. By extending the ontology, these observations can be generalized to other scholarly entities, e.g., articles or institutions, for the prediction of other scholarly patterns, e.g., co-citations or academic collaboration.     
\end{abstract}
\section{Introduction}
Knowledge semantically represented in knowledge graphs can be exploited to solve a broad range of problems in the respective domain. For example, in scientific domains, such as bio-medicine, scholarly communication, or even in industries, knowledge graphs enable not only the description of the meaning of data, but the integration of data from heterogeneous sources and the discovery of previously unknown patterns. 
With the rapid growth in the number of publications, scientific groups, and research topics, the availability of scholarly datasets has considerably increased. This generates a great challenge for researchers, particularly, to keep track of new published scientific results and potential future co-authors.  
To alleviate the impact of the explosion of scholarly data, knowledge graphs provide a formal framework where 
scholarly datasets can be integrated and diverse knowledge-driven tasks can be addressed. 
Nevertheless, to exploit the semantics encoded in such knowledge graphs, a deep analysis of the graph structure as well as the semantics of the represented relations, is required. 
There have been several attempts considering both of these aspects.
However, the majority of previous approaches rely on the topology of the graphs and usually omit the encoded meaning of the data.
Most of such approaches are also mainly applied on special graph topologies, e.g., ego networks rather than general knowledge graphs.
To provide an effective solution to the problem of representing scholarly data in knowledge graphs, and exploiting them to effectively support knowledge-driven tasks such as pattern discovery, we propose \koronaB, a knowledge-driven framework for scholarly knowledge graphs. \koronaB enables both the creation of scholarly knowledge graphs and knowledge discovery. 
Specifically, \koronaB resorts to community detection methods and semantic similarity measures to discover hidden relations in scholarly knowledge graphs. 
We have empirically evaluated the performance of \koronaB in a knowledge graph of publications and researchers from the Semantic Web area. As a proof of concept, we studied the accuracy of identifying co-author networks. Further, the predictive capacity of \koronaB has been analyzed by members of the Semantic Web area. 
Experimental outcomes suggest the next conclusions: 
\begin{inparaenum}[\itshape i\upshape)]
\item \koronaB identifies co-author networks that include researchers that both work on similar topics, and attend and publish in the same scientific venues.   
\item \koronaB allows for uncovering scientific relations among researchers of the Semantic Web area. 
\end{inparaenum}
The contributions of this paper are as follows:
\begin{itemize}
\item A scholarly knowledge graph integrating data from DBLP datasets;
\item The \koronaB knowledge-driven framework, which has been implemented on top of two graph partitioning tools, semEP~\cite{Palma2014} and METIS~\cite{karypis1998fast}, and relies on semantic similarity to identify patterns in a scholarly knowledge graph; 
\item Collaboration suggestions based on co-author networks; and 
\item An empirical evaluation of the quality of \koronaB using semEP and METIS.  
\end{itemize}

This paper includes five additional sections. 
Section~\ref{sec:motivating}  motivates our work with an example.
The \koronaB approach is presented in section~\ref{sec:approach}.
Related work is analyzed in section~\ref{sec:experimental}.
Section~\ref{sec:relatedWork} reports on experimental results.
Finally, section~\ref{sec:conclusions} concludes and presents ideas for future work.

\section{Motivating Example}
\label{sec:motivating}
\begin{figure}[t]
\vspace*{-.3cm}
    \subfloat[Researchers working on similar topics were in two co-authorship communities.]{
    \label{fig:motiv1}
    \includegraphics[scale=0.48]{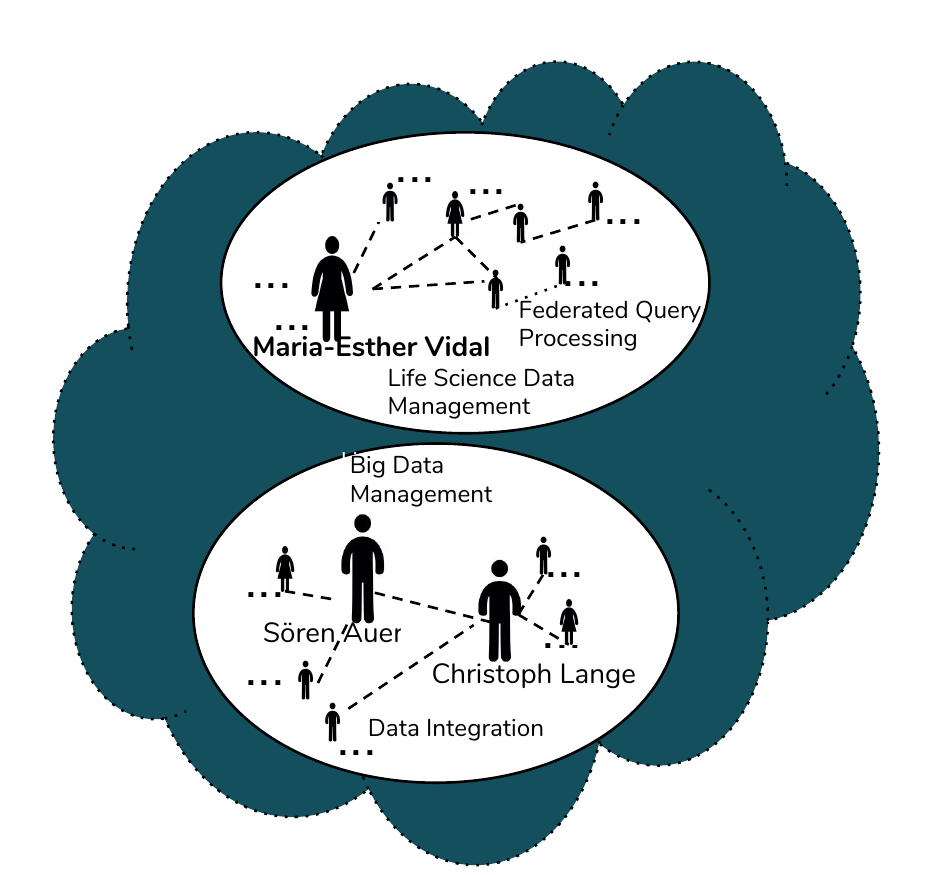}
    }
\qquad
\subfloat[Researchers working on similar topics constitute a co-authorship community and produce a large number of scholarly artifacts.]{
    \label{fig:motiv2}
    \includegraphics[scale=0.48]{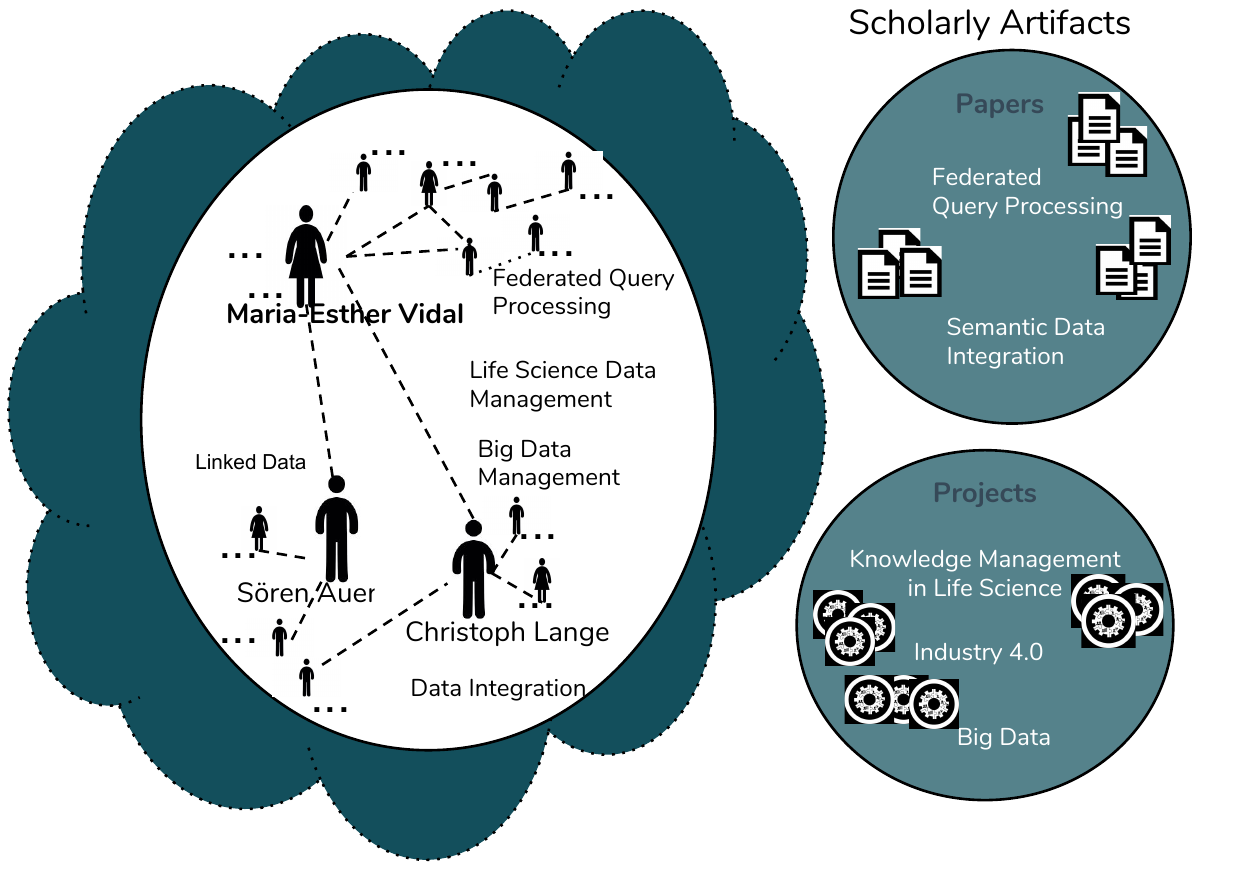}
    }
    \caption{{\bf Motivating Example}. Co-authorship communities from the Semantic Web area working on data-centric problems. Researchers were in different co-authorship communities (2016)  (a) started a successful scientific collaboration in 2016 (b), and as a result, produced a large number of scholarly artifacts.}
    \vspace*{-.1cm}
    \end{figure}
In this section, we motivate the problem of knowledge discovery tackled in this paper.
We present an example of co-authorship relation discovery between researchers working on data-centric problems in the Semantic Web area.
We checked the Google Scholar profiles of three researchers between 2015 and 2017, and compared their networks of co-authorship. By 2016, S\"{o}ren Auer and Christoph Lange were part of the same research group and wrote a large number of joint publications. Similarly, Maria-Esther Vidal, also working on data management topics, was part of a co-authorship community. Figure\autoref{fig:motiv2} illustrates the two co-authorship communities, which were confirmed by the three researchers. After 2016, these three researchers started to work in the same research lab, and a large number of scientific results, e.g., papers and projects, was produced. An approach able to discover such potential collaborations automatically would allow for the identification of the best collaborators and, thus, for maximizing the success chances of scholars and researchers working on similar scientific problems. In this paper, we rely on the natural intuition that successful researchers working on similar problems and producing similar solutions can collaborate successfully, and propose \korona, a framework able to discover unknown relations between scholarly entities in a knowledge graph. \koronaB implements graph partitioning methods able to exploit semantics encoded in a scholarly knowledge graph and to identify communities of scholarly entities that should be connected or related.     
 \section{Our Approach: \korona}
\label{sec:approach}
\subsection{Preliminaries}
\label{sec:preliminaries}
\begin{figure}[t!]
\vspace*{-.3cm}
    \centering
     \includegraphics[width=1\textwidth]{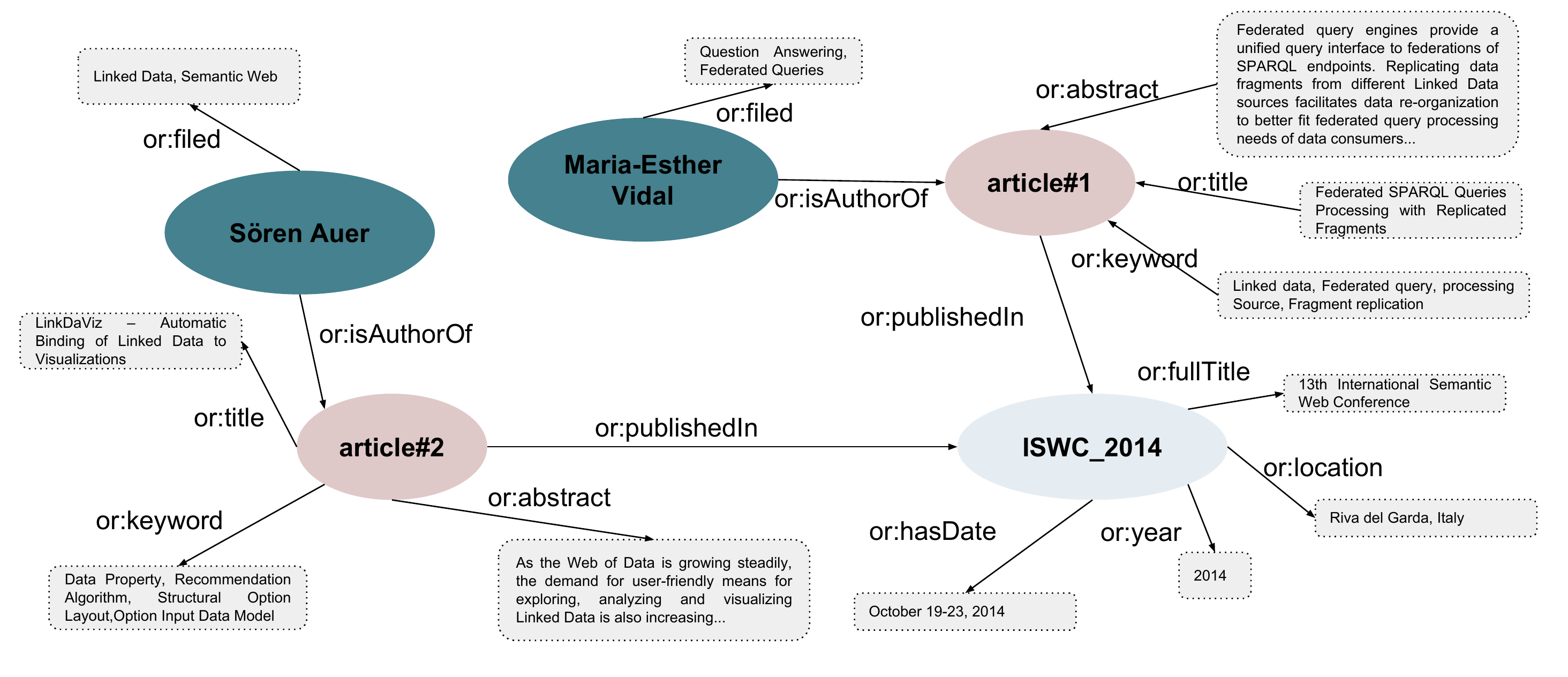}
    \caption{{\bf \koronaB Knowledge Graph.} Scholarly entities and relations.}
    \label{fig:kg}
    \vspace*{-.1cm}
\end{figure}

\begin{figure}[t!]
  \subfloat[Network of Researchers and Articles.]{
    \includegraphics[scale=0.41]{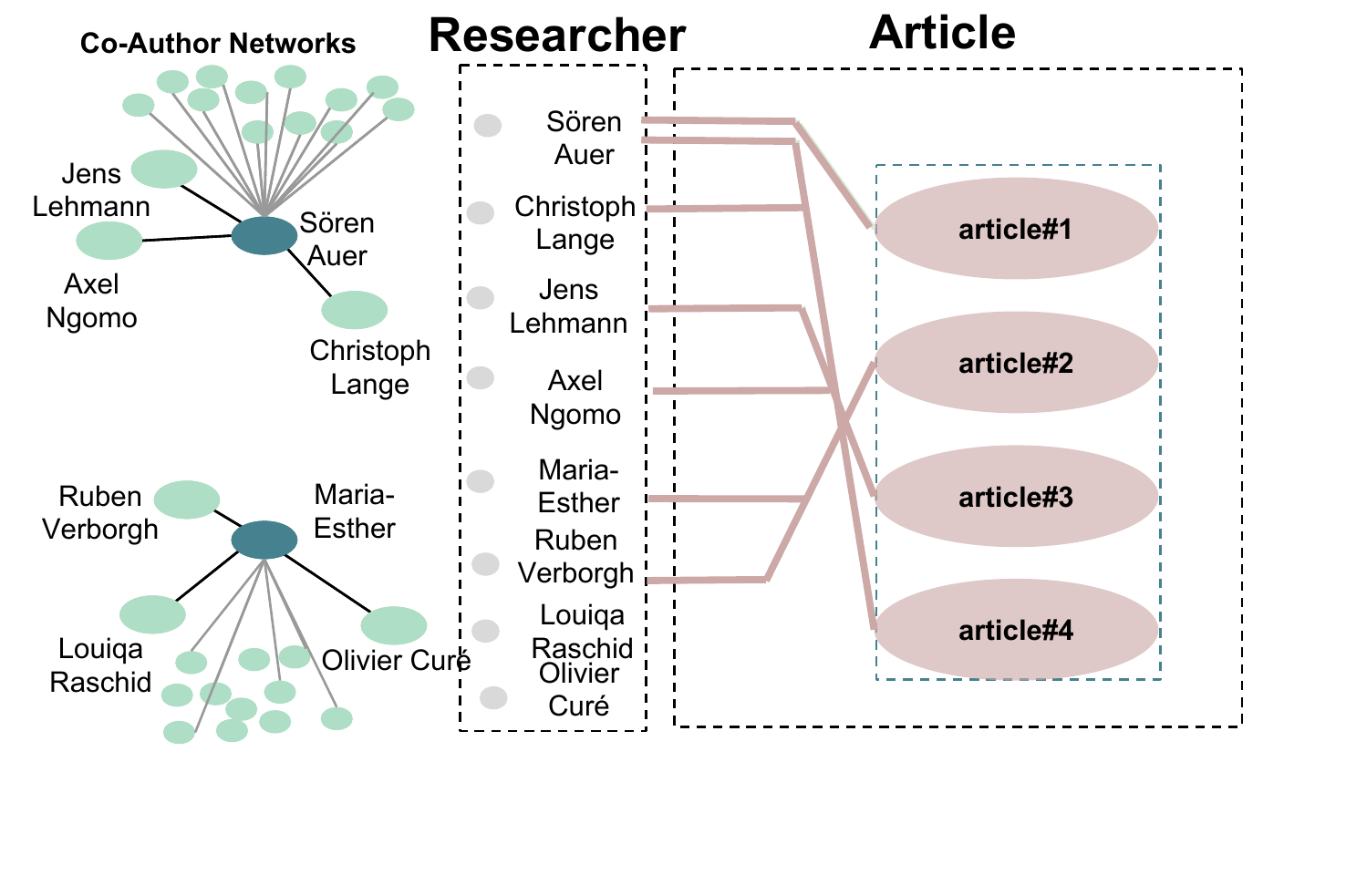}
    \label{fig:researchers}
  }
  \quad
  \subfloat[Networks of Events and Articles.]{
    \includegraphics[scale=0.41]{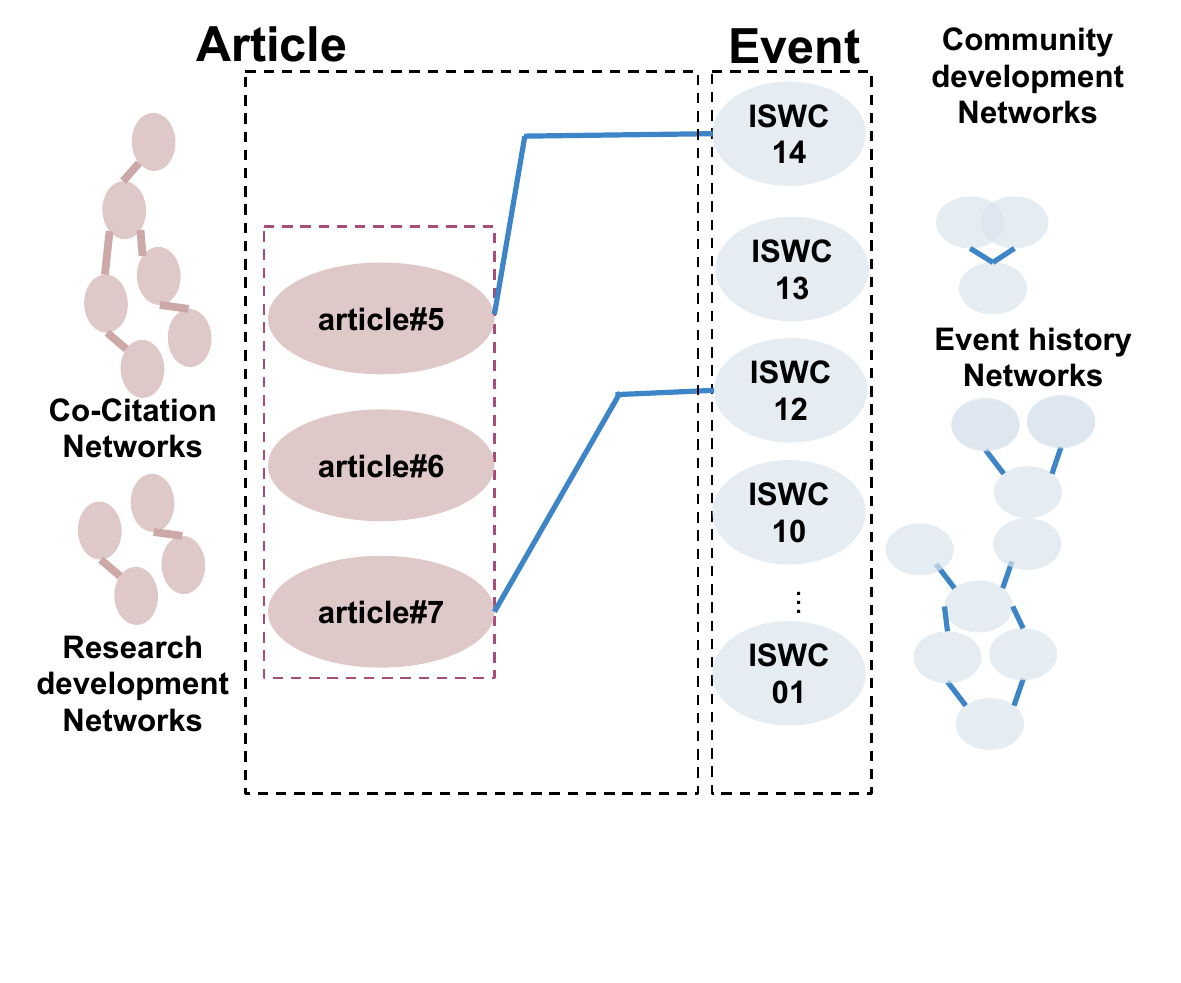}
    \label{fig:events}
  }
  \vspace*{-.1cm}
  \caption{{\bf Scholarly networks}. (a) Co-authors networks  from researchers and articles.(b) Co-citation networks from discovered from events and articles.}
  \label{fig:networks}
\end{figure}

The definitions required to understand our approach are presented in this section. First, we define a scholarly knowledge graph as a knowledge graph where nodes represent scholarly entities of different types, e.g., publications, researchers, publication venues, or scientific institutions, and edges correspond to an association between these entities, e.g., co-authors or citations.
\begin{mydef}{Scholarly Knowledge Graph.}
Let $U$ be a set of RDF URI references and $L$ a set of RDF literals.
Given sets $V_{e}$ and $V_{t}$ of scholarly entities and types, respectively, and given a set $P$ of properties representing scholarly relations, a scholarly knowledge graph is defined as $\cal{SKG}$=$(V_{e} \cup V_{t}, E,P)$, where:
\begin{itemize}
    \item Scholarly entities and types are represented as RDF URIs, i.e., $V_{e} \cup V_{t} \subseteq U$;
     \item Relations between scholarly entities and types are represented as RDF properties, i.e., $P \subseteq U$ and $E \subseteq (V_{e} \cup V_{t} \times P \times V_{e} \cup V_{t} \cup L)$
 \end{itemize}
 \end{mydef}
 Figure~\ref{fig:kg} shows a portion of a scholarly knowledge graph describing scholarly entities, e.g., papers, publication venues, researchers, and different relations among them, e.g., co-authorship, citation, and collaboration. 
  \begin{mydef}{Co-author Network.}
 A co-author network $\cal{CAN}$=$(V_{a},E_{a},P_{a})$ corresponds to a subgraph of $\cal{SKG}$=$(V_{e} \cup V_{t}, E,P)$, where
      \begin{itemize}
     \item Nodes are scholarly entities of type \emph{researcher},  
     \[V_{a}=\{a \mid (a\; \textit{rdf:type} \; \textit{:Researcher}) \in E\}\]
     \item Researchers are related according to co-authorship of scientific publications,
     \begin{nospaceflalign*}
    E_{a}=\{&(a_i \; \textit{:co-author} \; a_j) \mid \exists p\; .\; a_i, a_j \in V_{a} \; \wedge \;(a_i\; \textit{:author} \;p ) \in E \; \wedge\; & \\
    &(a_j\; \textit{:author} \;p ) \in E \; \wedge (p\; \textit{rdf:type} \; \textit{:Publication}) \in E \}
    \end{nospaceflalign*}
     \end{itemize}
\end{mydef}
\autoref{fig:networks} shows scholarly networks that can be generated by \korona.
Some of these networks are among the recommended applications for scholarly data analytics in~\cite{XiaWBL17}.
However, the focus on this work is on co-author networks. 

\subsection{Problem Statement}
Let $\cal{SKG}'$=$(V_{e} \cup V_{t}, E',P)$ and $\cal{SKG}$=$(V_{e} \cup V_{t}, E,P)$ be two scholarly knowledge graphs, such that $\cal{SKG}'$ is an \emph{ideal} scholarly knowledge graph that contains all the \emph{existing and successful relations} between scholarly entities in $V_{e}$, i.e., an oracle that knows whether two scholarly entities should be related or not. $\cal{SKG}$=$(V_{e} \cup V_{t}, E,P)$ is the \emph{actual} scholarly knowledge graph, which only contains a portion of the relations represented in  $\cal{SKG}'$, i.e., $E \subseteq E'$; it represents those relations that are known and is not necessarily complete. Let $\Delta(E', E) = E' - E$ be the set of relations existing in the ideal scholarly knowledge graph $\cal{SKG}'$ that are not represented in the actual scholarly knowledge graph $\cal{SKG}$. Let $\cal{SKG}_\text{comp}$=$(V_{e} \cup V_{t}, E_\text{comp},P)$
be a \emph{complete} knowledge graph, which includes a relation for each possible combination of scholarly entities in $V_{e}$ and properties in $P$, i.e., 
$E\subseteq E'\subseteq E_\text{comp}$. Given a relation $e \in \Delta(E_\text{comp}, E)$, the problem of discovering scholarly relations consists in determining whether $e \in E'$, i.e., whether a relation $r$=$(e_i \; p \; e_j)$ corresponds to an existing relation in the ideal scholarly knowledge graph $\cal{SKG}'$.

In this paper, we specifically focus on the problem of discovering \textit{successful co-authorship relations} between researchers in scholarly knowledge graph $\cal{SKG}$=$(V_{e} \cup V_{t}, E,P)$. Thus, we are interested in finding the co-author network $\cal{CAN}$=$(V_{a},E_{a},P_{a})$ composed of the maximal set of relationships
or edges that belong to the ideal scholarly knowledge graph, i.e., the set $E_{a}$ in $\cal{CAN}$ that corresponds to a solution of the following optimization problem:
\begin{equation}
\label{equa1}
 \argmax_{\mathit{E_{a}}\subseteq \mathit{E_{comp}}}{|E_{a} \cap E'|}
 \vspace*{-.5cm}
 \end{equation}
\subsection{Proposed Solution}
We propose \koronaB to solve the problem of discovering meaningful co-authorship relations between researchers in scholarly knowledge graphs. \koronaB relies on information about relatedness between researchers to identify communities composed of researchers that work on similar problems and publish in similar scientific events. \koronaB is implemented as an unsupervised machine learning method able to partition a scholarly knowledge graph into subgraphs or communities of co-author networks. Moreover, \koronaB applies the \emph{homophily} prediction principle over the communities of co-author networks to identify successful co-author relations between researchers in the knowledge graph. The \emph{homophily} prediction principle states that similar entities tend to be related to similar entities~\cite{liben2007link}. Intuitively, the application of the \emph{homophily} prediction principle enables \koronaB to relate two researchers $r_i$ and $r_j$ whenever they work on similar research topics or publish in similar scientific venues. The relatedness or similarity between two scholarly entities, e.g., researchers, research topics, or scientific venues, is represented as RDF properties in the scholarly knowledge graph. Semantic similarly measures, e.g., GADES~\cite{RibonVKS16} or Doc2Vec~\cite{Le2014}, are utilized to quantify the degree of relatedness between two scholarly entities. 
The identified degree shows the relevance of entities and returns the most related ones.

\begin{figure}[t]
  \centering
    \includegraphics[width=0.9\textwidth]{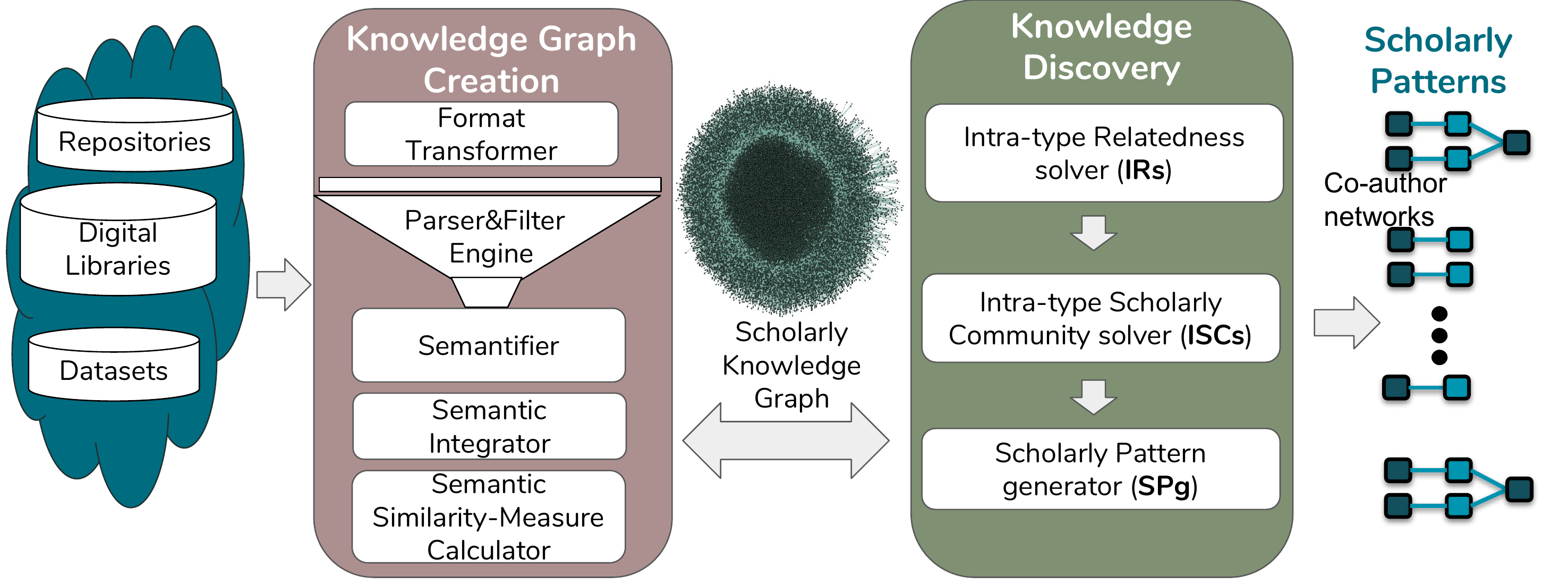}
 \caption{\textbf{The \koronaB Architecture}. \koronaB receives scholarly datasets and outputs scholarly patterns, e.g., co-author networks. First, a scholarly knowledge graph is created. Then, community detection methods and similarity measures are used to compute communities of scholarly entities and scholarly patterns. 
 }
  \label{fig:korona}
 \end{figure}
\begin{figure}[h!]
  \subfloat[Similarity-based Relatedness]{
    \includegraphics[scale=0.43]{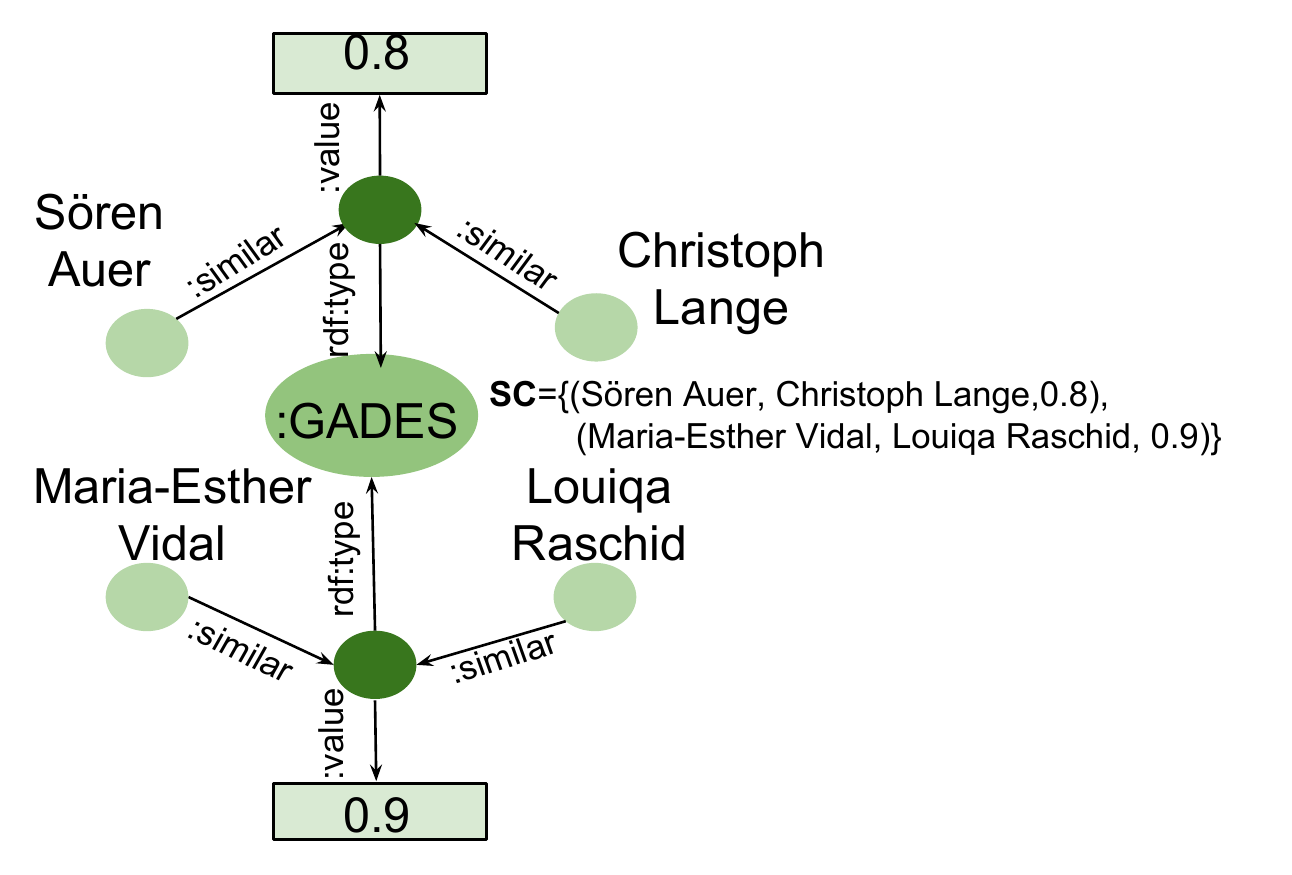}
    \label{fig:sim1}
  }
  \quad
  \subfloat[Path-based Relatedness]{
    \includegraphics[scale=0.43]{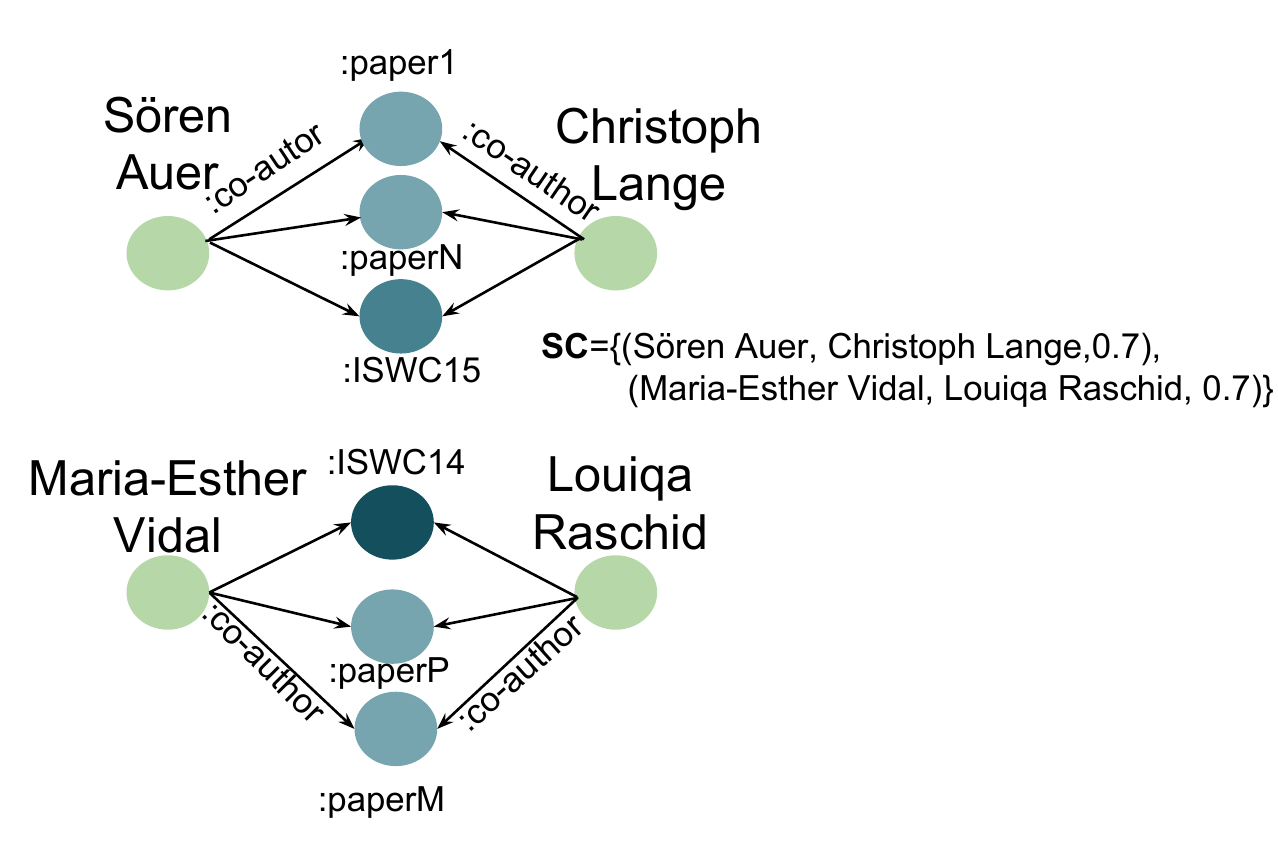}
    \label{fig:sim2}
  }
  \caption{{\bf Intra-type Relatedness solver (\textbf{IRs})}. Relatedness across scholarly entities. (a) Relatedness is computed according to the values of a semantic similarity metrics, e.g., GADES. (b) Relatedness is determined based on the number of paths between two scholarly entities.}
  \label{fig:sim}
 \vspace*{-.2cm}
\end{figure}
\begin{figure}[t!]
\vspace*{-.5cm}
    \centering
     \includegraphics[width=0.85\textwidth]{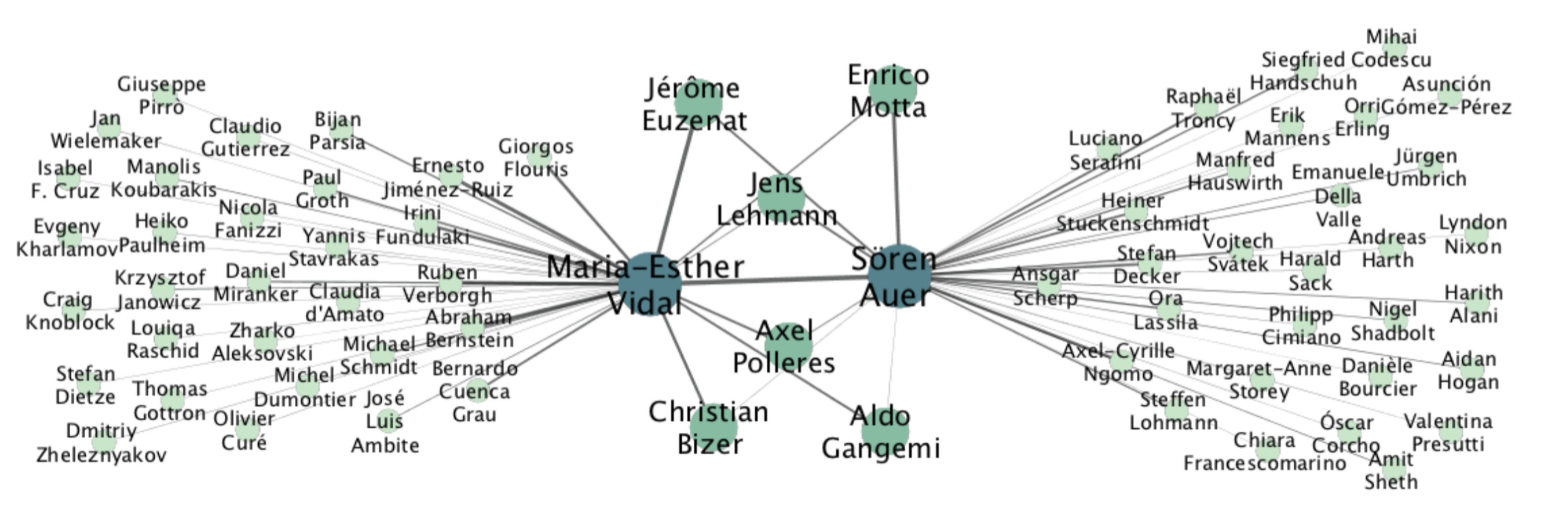}
    \caption{{\bf Co-author network.} A network generated from scholarly communities.}
    \label{fig:conetwork}
    \vspace*{-.2cm}
\end{figure}

Figure~\ref{fig:korona} depicts the \koronaB architecture; it implements a knowledge-driven approach able to transform scholarly data ingested from publicly available data sources into patterns that represent discovered relationships between researchers. Thus, \koronaB receives scholarly data sources and outputs co-author networks; it works in two stages:   
\begin{inparaenum}[\it (a)]
\item Knowledge graph creation and 
\item Knowledge graph discovery.
\end{inparaenum} 
During the knowledge graph creation stage, a semantic integration pipeline is followed in order to create a scholarly knowledge graph from data ingested from heterogeneous scholarly data sources.  
It utilizes mapping rules between the \koronaB ontology and the input data sources to create the scholarly knowledge graph. Additionally, semantic similarity measures are used to compute the relatedness between scholarly entities; the results are explicitly represented in the knowledge graph as scores in the range of 0.0 and 1.0. The knowledge graph creation stage is executed offline and enables the integration of new entities in the knowledge graph whenever the input data sources change. 
On the other hand, the knowledge graph discovery step is executed \emph{on the fly} over an existing scholarly knowledge graph. During this stage, \koronaB executes three main tasks: 
\begin{inparaenum}[\it (i)]
\item Intra-type Relatedness solver (\textbf{IRs}); 
\item Intra-type Scholarly Community solver (\textbf{IRSCs});
and 
\item Scholarly Pattern generator (\textbf{SPg}).
\end{inparaenum} 
\begin{figure}[t!]
\vspace*{-.4cm}
  \subfloat[Relatedness Across Researchers]{
    \includegraphics[scale=0.26]{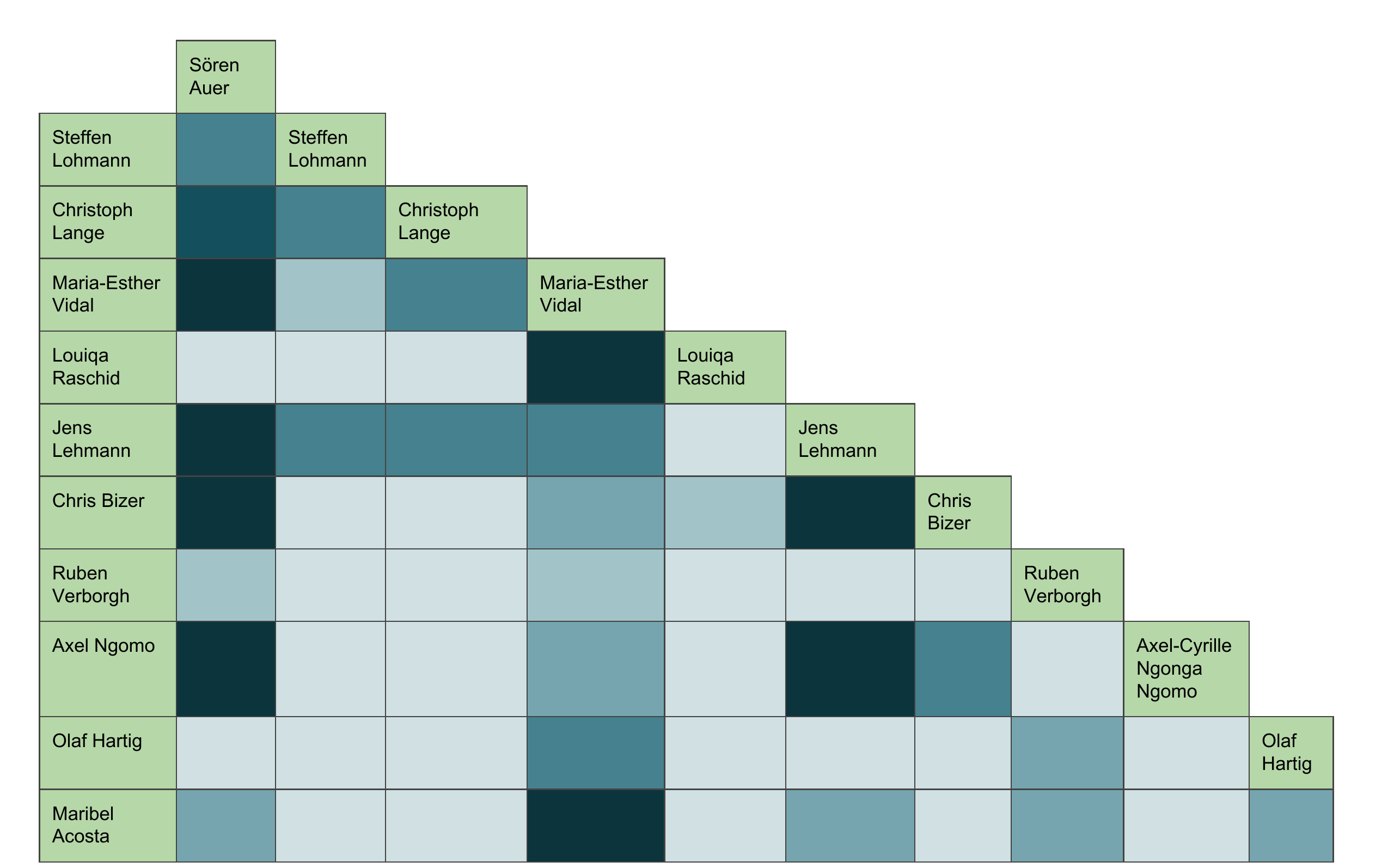}
    \label{fig:sim3}
  }
  \quad
  \subfloat[Communities of Researchers]{
    \includegraphics[scale=0.38]{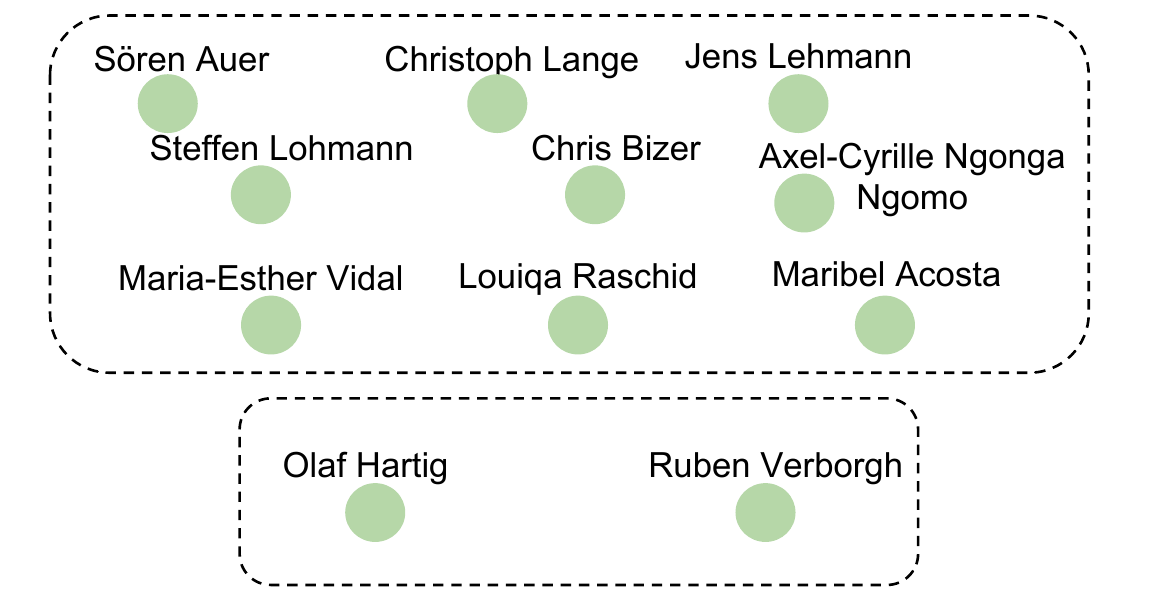}
    \label{fig:com}
  }
  \caption{{\bf Intra-type Relatedness solver (\textbf{IRs})}. Communities of similar researchers are computed. (a) The tabular representation of $\cal{SC}$; lower and higher values of similarity are represented by lighter and darker colors, respectively. (b) Two communities of researchers; each one includes highly similar researchers.}
  \label{fig:com1}
  \vspace*{-.2cm}
\end{figure}
\vspace*{-.4cm}
\paragraph{\bf Intra-type Relatedness solver (\textbf{IRs}).}
This module quantifies relatedness between the scholarly entities of the same type in a scholarly knowledge graph $\cal{SKG}$=$(V_{e} \cup V_{t}, E,P)$. \textbf{IRs} receives as input $\cal{SKG}$=$(V_{e} \cup V_{t}, E,P)$ and a scholarly type $V_a$ in $V_{t}$; it outputs a set $\cal{SC}$ of triples $(e_i,e_j,\textit{score})$, where $e_i$ and $e_j$ belong to $V_a$ and \emph{score} quantifies the relatedness between $e_i$ and $e_j$. The relatedness can be just computed in terms of the values of similarity represented in the knowledge graph, e.g., according to the values of the semantic similarity according to GADES or Doc2Vec. Alternatively, the values of relatedness can be computed based on the number of paths in the scholarly knowledge graph that connect the scholarly entities $e_i$ and $e_j$. \autoref{fig:sim} depicts two representations of the relatedness of scholarly entities. As shown in Figure\autoref{fig:sim1}, \textbf{IRs} generates a set $\cal{SC}$ according to the GADES values of semantic similarity; thus, \textbf{IRs} includes two triples $(\text{S\"{o}ren Auer}, \text{Christoph Lange},0.8)$, $(\text{Maria-Esther Vidal}, \text{Louiqa Raschid},0.9)$ in $\cal{SC}$. On the other hand, if paths between scholarly entities are considered (Figure\autoref{fig:sim2}), the values of relatedness can different, e.g., in this case,  S\"{o}ren Auer and Christoph Lange are equally similar as Maria-Esther Vidal and Louiqa Raschid.
 \vspace*{-.4cm}
\paragraph{\bf Intra-type Scholarly Community solver (\textbf{IRSCs}).}
Once the relatedness between the scholarly entities has been computed, communities of highly related scholarly entities are determined. \textbf{IRSCs} resorts to unsupervised methods such as METIS or semEP, and to relatedness values stored in $\cal{SC}$, to compute the scholarly communities. \autoref{fig:com1} depicts scholarly communities computed by \textbf{IRSCs} based on similarity values; as observed, each community includes researchers that are highly related; for readability, $\cal{SC}$ is shown as a heatmap where lower and higher values of similarity are represented by lighter and darker colors, respectively. For example, in Figure\autoref{fig:sim3}, S\"{o}ren Auer, Christoph Lange, and Maria-Esther Vidal are quite similar, and they are in the same community. 
\vspace*{-.4cm}
\paragraph{\bf Scholarly Pattern generator (\textbf{SPg}).}
\textbf{SPg} receives communities of scholarly entities and produces a network, e.g., a co-author network. \textbf{SPg} applies the {\it homophily} prediction principle on the input communities, and connects the scholarly entities in one community in a network. \autoref{fig:conetwork} shows a co-author network computed based on a scholarly knowledge graph created from DBLP; as observed, S\"{o}ren Auer, Christoph Lange, and Maria-Esther Vidal are included in the same co-author network. In addition to computing the scholarly networks, \textbf{SPg} scores the relations in a network and computes the \emph{weight of connectivity} of a relation between two entities. For example, in \autoref{fig:conetwork}, thicker lines represent strongly connected researchers in the network. \textbf{SPg} can also filter from a network the relations labeled  with higher values of weight of connectivity. All the relations in a network correspond to solutions to the problem of discovering \emph{successful co-authorship relations} defined in \autoref{equa1}.  To compute the weights of connectivity, \textbf{SPg} considers the values of similarity of the scholarly entities in a community $C$; weights are computed as aggregated values using an aggregation function $f(.)$, e.g., average or triangular norm. For each pair $(e_i,e_j)$ of scholarly entities in $C$, the weight of connectivity  between $e_i$ and $e_j$,  $\phi(e_i,e_j\mid C)$, is defined as: $\phi(e_i,e_j\mid C)= \{f(\textit{score})\mid e_z, e_q \in C \wedge  (e_z,e_q,\textit{score}) \in \cal{SC}\}$.
\section{Empirical Evaluation}
\label{sec:experimental}
\subsection{Knowledge Graph Creation}
\label{sec:kgc}
A scholarly knowledge graph has been crafted using the DBLP collection (7.83 GB in April 2017\footnote{\url{http://dblp2.uni-trier.de/e55477e3eda3bfd402faefd37c7a8d62/}}); it includes researchers, papers, and publication year from the International Semantic Web Conference (ISWC) 2001–2016. The knowledge graph also includes similarity values between researchers who have published at ISWC (2001–2017).  Let $\mathit{PC}_{e_i}$ and $\mathit{PC}_{e_j}$ be the number of papers published by researchers $e_i$ and $e_j$ together (as co-authors), respectively at ISWC (2001–2017). Let $\mathit{TP}_{e_i}$ and $\mathit{TP}_{e_j}$ be the total number of papers that $e_i$ and $e_j$ have in all conferences of the scholarly knowledge graph, respectively. The similarity measure is defined as:   $\mathit{SimR}(e_i, e_j) = \frac{ \mathit{PC}_{e_i} \cap \mathit{PC}_{e_j}}{ \mathit{TP}_{e_i} \cup \mathit{TP}_{e_j} }$. The similarities between ISWC (2002–2016) are represented as well. Let $\mathit{RC}_i$ and $\mathit{RC}_j$ the number of the authors with papers published in conferences $c_i$ and $c_j$ respectively. The similarity measure corresponds to $SimC(c_i, c_j) = \frac{ RC_i \cap RC_j}{ RC_i \cup RC_j}$.
Thus, the scholarly knowledge graph includes both scholarly entities enriched with their values of similarity.
\begin{figure}[t!]
\vspace*{-.3cm}
		\centering
		\small
		\begin{tabular}{cc}
		{\frame{\includegraphics[width=0.4\linewidth]{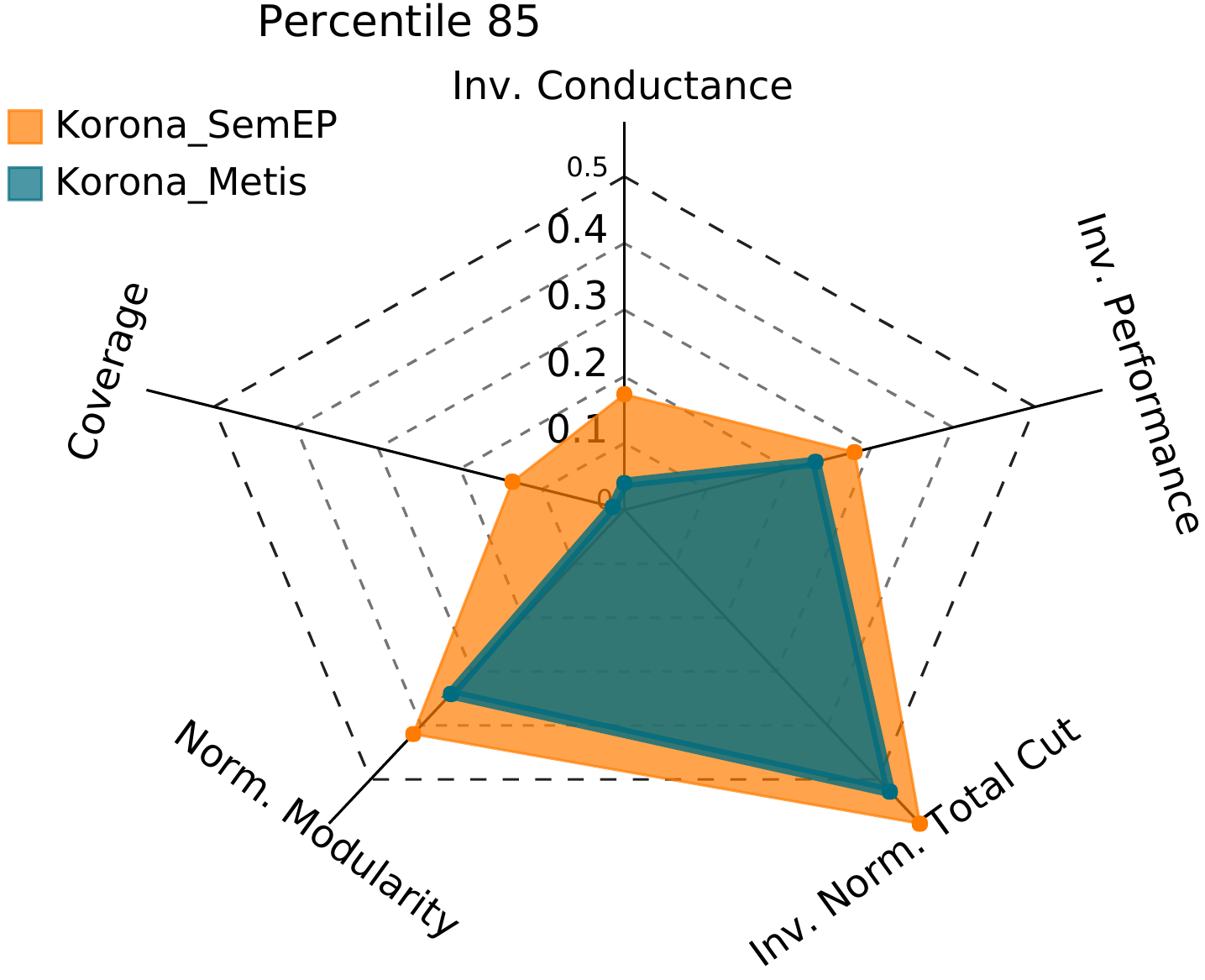}}}& {\frame{\includegraphics[width=0.4\linewidth]{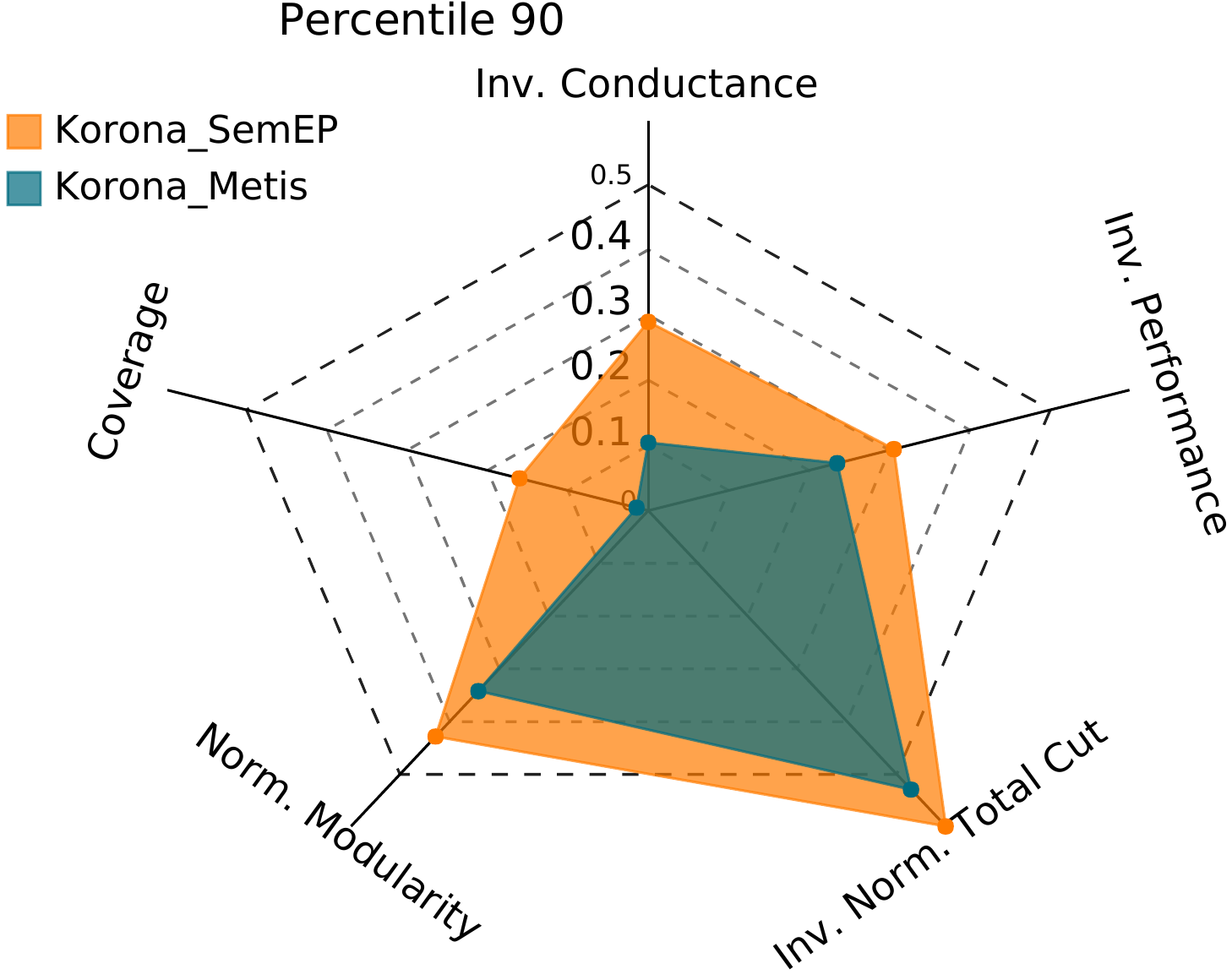}}}\\
		{(a) Percentile 85}&
		{(b) Percentile 90}\\
		{\frame{\includegraphics[width=0.4\linewidth]{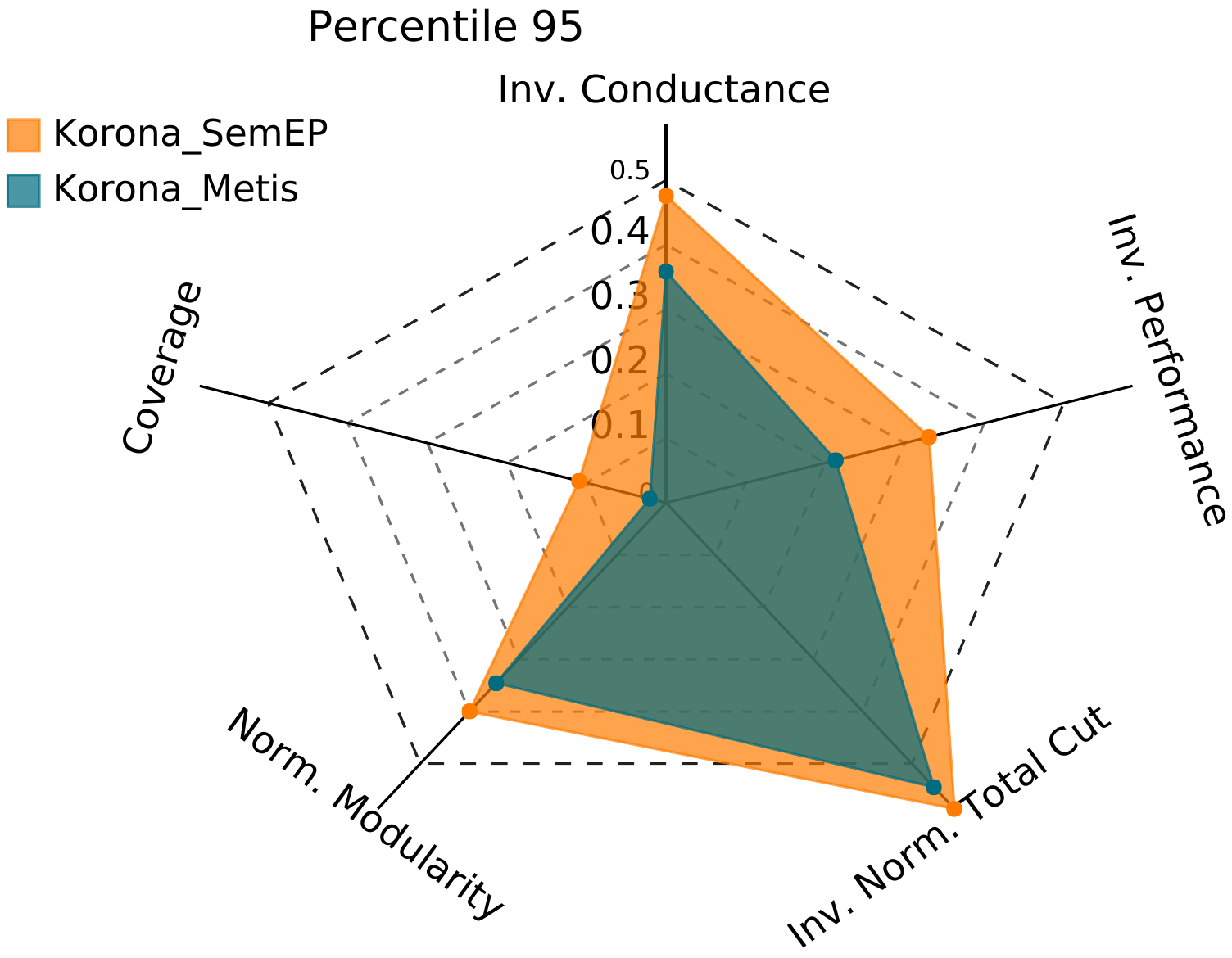}}}&
		{\frame{\includegraphics[width=0.4\linewidth]{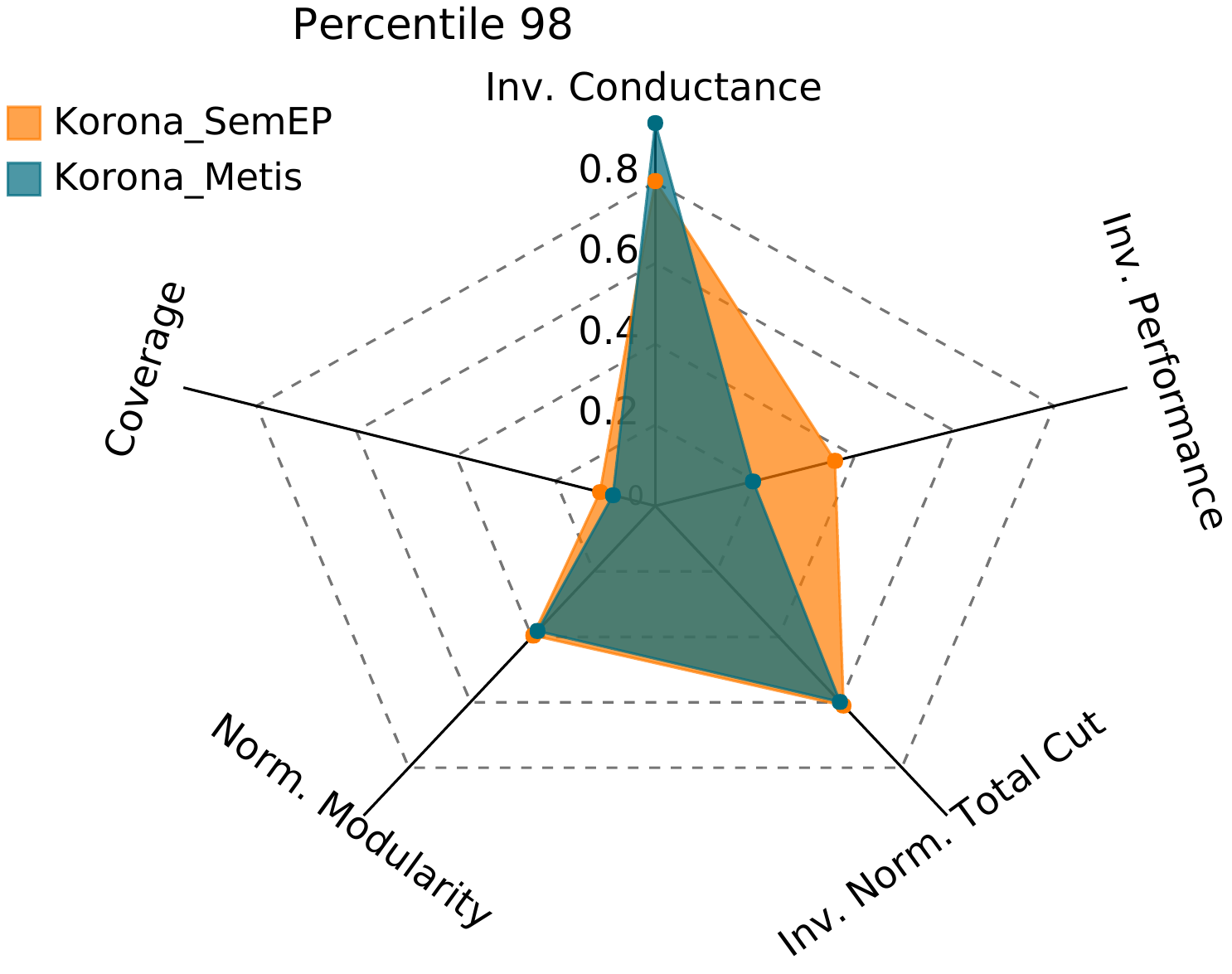}}}\\
        {(c) Percentile 95}&{(d) Percentile 98}
		\end{tabular}
\caption{{\bf Quality of \korona.} Communities evaluated in terms of prediction metrics (higher values are better); percentiles
    85, 90, 95, and 98 are reported. \koronaB exhibits the best performance at percentile 95 and groups  similar researchers according to research topics and events where they publish. }
\label{fig:iter1}
\vspace*{-.1cm}
\end{figure}
\subsection{Experimental Study}
The effectiveness of \koronaB has been evaluated in terms of the quality of both the generated communities of researchers and the predicted co-author networks.
\vspace*{-.6cm}
\paragraph{\bf Research Questions:} 
We assess the following research questions:
\begin{inparaenum}[\bf {\bf RQ}1\upshape)]
\item Does the semantics encoded in scholarly knowledge graphs impact the quality of scholarly patterns?
\item Does the semantics encoded in scholarly knowledge graph allow for improving the quality of the predicting co-author relations?
\end{inparaenum}
\vspace*{-.3cm}
\paragraph{\bf Implementation:} \koronaB is implemented in Python 2.7. The experiments were executed on a macOS High Sierra 10.13 (64 bits) Apple MacBook Air machine with an Intel Core i5 1.6 GHz CPU and 8 GB RAM. METIS 5.1~\footnote{\url{http://glaros.dtc.umn.edu/gkhome/metis/metis/download}} and
SemEP~\footnote{\url{https://github.com/gpalma/semEP}} are part of \koronaB and used to obtain the scholarly patterns.
\vspace*{-.3cm}
\paragraph{\bf Evaluation metrics:}
Let $Q = \{C_1, \dots C_n\}$ be the set of communities obtained by \korona: 
\textit{Conductance}: measures relatedness of entities in a community, and how different they are to entities outside the community~\cite{Gaertler2005}. The inverse of the conductance $1 - \mathit{Conductance}(S)$ is reported.
\textit{Coverage}: compares the fraction of intra-community similarities among entities to the sum of all similarities among entities~\cite{Gaertler2005}. 
\textit{Modularity}: is the value of the intra-community similarities among the entities divided by the sum of all the similarities among the entities, minus the sum of the similarities among the entities in different communities, in the case they were randomly distributed in the communities~\cite{newman2006modularity}.
The value of the modularity lies in the range $[-0.5 , 1 ]$, which can be scaled to $[0, 1]$ by computing $\frac {\mathit{Modularity}(Q) + 0.5} {1.5}$. 
 \textit{Performance}: sums the number of intra-community relationships, plus the number of non-existent relationships between communities~\cite{Gaertler2005}. 
 \textit{Total Cut}: sums all similarities among entities in different communities~\cite{Bulu2016}. Values of total cut are normalized by dividing by the sum of the similarities among the entities; inverse values are reported, i.e., $1-\mathit{NormTotalCut}(Q)$.
\paragraph{\bf Experiment 1: Evaluation of the Quality of Collaboration Patterns.}
Prediction metrics are used to evaluate the quality of the communities generated by \koronaB using METIS and semEP; relatedness of the researchers is measured in terms of $SimR$ and $SimC$. Communities are built according to different similarity criteria; percentiles of 85, 90, 95, and 98 of the values of similarity are analyzed. For example, in percentile 85 only 85\% of all similarity values among entities have scores lower than the similarity value in the percentile 85. Figure~\ref{fig:iter1} presents the results of the studied metrics. In general, in all percentiles, the communities include closely related researchers. However, both implementations of \koronaB exhibit quite good performance at percentile 95, and allow for grouping together researchers that are highly related in terms of the research topics on which they work, and the events where their papers are published. On the contrary, \koronaB creates many communities of no related authors for percentiles 85 and 90, thus exposing low values of coverage and conductance.   
\vspace*{-.3cm}
\paragraph{\bf Experiment 2: Survey of the Quality of the Prediction of Collaborations among Researchers.}
\begin{table*}[t!]
\scriptsize
\centering
\begin{tabular}{p{12cm}r}
\toprule
\textbf{Q1. Do you know this person? Have you co-authored before?} To avoid confusion, the meaning of ``knowing'' was kept simple and general. 
The participants were asked to only consider if they were aware of the existence of the recommended person in their research community. \\ 
\midrule
\textbf{Q2. Have you co-authored ``before'' with this person at any event of the ISWC series?}
With the same intent of keeping the survey simple, all types of collaboration on papers in any edition of this event series were considered as ``having co-authored before''.\\ 
\midrule
\textbf{Q3. Have you co-authored with this person after May 2016?}
Our study considered scholarly metadata of publications until May 2016.
The objective of this question was to find out whether a prediction had actually come true, and the researchers had collaborated.\\ 
\midrule
\textbf{Q4. Have you ever planned to write a paper with the recommended person and you never made it and why?}
The aim is to know whether two researchers who had been predicted to work together actually wanted to but then did not and the reason, e.g., geographical distance.\\ 
\midrule
\textbf{Q5. On a scale from 1–5, (5 being most likely), how do you score the relevance of your research with this person?}
The aim is to discover how close and relevant are the collaboration recommendations to the survey participant.
\\ 
  \bottomrule
\end{tabular}
\vspace*{-.2cm}
\caption{{ \bf Survey}. Questions to validate the recommended collaborations.}
\label{tab:questions}
\end{table*}
Results of an online survey\footnote{\url{https://bit.ly/2ENEg2G}} among 10 researchers are reported;
half of the researchers are from the same research area, while the other half was chosen randomly.
Knowledge subgraphs of each of the participants are part of the \koronaB research knowledge graph; predictions are computed from these subgraphs.
The predictions for each were laid out in an online spreadsheet along with 5 questions and a comment section.
\autoref{tab:questions} lists the five questions that the survey participants were asked to validate the answers, while \autoref{tab:results} reports on the results of the study. The analysis of results suggests that \koronaB predictions represent potentially \emph{successful co-authorship relations}; thus, they provide a solution to the problem tackled in this paper.   

\begin{table*}[t!]
\scriptsize
\centering
\begin{tabular}{p{2.0cm}p{0.75cm}p{1.4cm}p{1.4cm}p{1.4cm}p{1.4cm}p{1.4cm}l}
\toprule
\textbf{\korona} & \% & Q.1(a) & Q.1(b) & Q.2 & Q.3 & Q.4 & Q.5\\
\midrule
Korona-METIS &	85	& 	0.26$\pm$0.25 &	0.72$\pm$0.29	&	0.99$\pm$0.04	&	0.86$\pm$0.13 & 0.86$\pm$0.20	&	\cellcolor{blue!25}3.10$\pm$0.59	\\
Korona-semEP &	85	& 0.24$\pm$0.21	& 0.80$\pm$0.34 &	1.00$\pm$0.00	&	0.97$\pm$0.07	&	0.93$\pm$0.16	& 	\cellcolor{blue!25}3.35$\pm$0.85	\\
\midrule
Korona-METIS &	90	& 0.39$\pm$0.24	&	0.91$\pm$0.19 	&	1.00$\pm$0.00	& 1.00$\pm$0.00 & 0.98$\pm$0.04	& \cellcolor{blue!25} 3.03$\pm$0.79	 	\\
Korona-semEP &	90	&	0.13$\pm$0.18	&	0.89$\pm$0.18	& 1.00$\pm$0.00		&	1.00$\pm$0.00	&	0.85$\pm$0.23	&		\cellcolor{blue!25}	3.12$\pm$1.06\\
\midrule
Korona-METIS &	95	& 0.40$\pm$0.34	&	0.93$\pm$0.08	&1.00$\pm$0.00	&	0.80$\pm$0.45	&0.95$\pm$0.10 	&	\cellcolor{blue!25}	3.20$\pm$0.81	\\
Korona-semEP &	95	& 0.14$\pm$0.30 & 0.81$\pm$0.40	&	0.67$\pm$0.58	&  0.60$\pm$0.55 & 0.69$\pm$0.47	&		\cellcolor{blue!25}	3.83$\pm$0.76\\
  \bottomrule
\end{tabular}
\vspace*{-.1cm}
\caption{{\bf Survey results.} Aggregated normalized values of negative answers provided by the study participants during the validation of the recommended collaborations (Q.1(a), Q.1(b), Q.2, Q.3, and Q.4); 
average (lower is better) and standard deviation (lower is better) are reported.  For Q.5, average and standard deviation of the scale from 1–5 are presented; higher average values are better.} 
\label{tab:results}
\end{table*}

\section{Related Work}
\label{sec:relatedWork}
Xia et al.~\cite{XiaWBL17} provides a comprehensive survey of tools and technologies for scholarly data management, as well as a review of data analysis techniques, e.g., social networks and statistical analysis.
However, all the proposals have been made over raw data and knowledge-driven methods were not considered. 
Wang et al.~\cite{wang2015link} present a comprehensive survey of link prediction in social networks, while Paulheim~\cite{paulheim2017knowledge} presents a survey of methodologies used for knowledge graph refinement; both works show the importance of the problem of knowledge discovery.
Traverso-Ribón et al.~\cite{KOI2016} introduces a relation discovery approach, \KOI, able to identify hidden links in TED talks; it relies on heterogeneous bipartite graphs and on the link discovery approach proposed in~\cite{Palma2014}.
In this work, Palma et al. present semEP, a semantic-based graph partitioning approach, which was used in the implementation of \korona-semEP. 
Graph partitioning of semEP is similar to \KOIB with the difference of only considering isolated entities, whereas \KOIB is desired for ego networks. 
However, it is only applied to ego networks, whereas \koronaB is mainly designed for knowledge graphs. 
Sachan and Ichise~\cite{sachan2010using} propose a syntactic approach considering dense subgraphs of a co-author network created from the DBLP. 
They discover relations between authors and propose pairs of researchers belonging to the same community. 
A link discovery tool is developed for the biomedical domain by Kastrin et al.~\cite{Kastrin2014}.
Albeit effective, these approaches focus on the graph structure and ignore the meaning of the data.

\section{Conclusions and Future Work}
\label{sec:conclusions}
\koronaB is presented for unveiling unknown relations; it relies on semantic similarity measures to discover hidden relations in scholarly knowledge graphs. 
Reported and validated experimental results show that \koronaB retrieves valuable information that can impact the research direction of a researcher. 
In the future, we plan to extend \koronaB to detect other networks, e.g., affiliation networks, co-citation networks and research development networks. 
 We plan to extend our evaluation over big scholarly datasets and study the scalability of \korona; further, the impact of several semantic similarity measures will be included in the study. Finally, \koronaB will be offered as an online service that will enable researchers to explore and analyze the underlying scholarly knowledge graph.
\vspace*{-.5cm}
\subsubsection*{Acknowledgement}
This work has been partially funded by the EU H2020 programme for the project iASiS (grant agreement No. 727658).
\vspace*{-.5cm}
\bibliographystyle{splncs04}
\bibliography{bibliography}

\begin{thebibliography}{10}
\providecommand{\url}[1]{\texttt{#1}}
\providecommand{\urlprefix}{URL }
\providecommand{\doi}[1]{https://doi.org/#1}

\bibitem{Bulu2016}
Bulu{\c{c}}, A., Meyerhenke, H., Safro, I., Sanders, P., Schulz, C.: Recent
  Advances in Graph Partitioning. Springer, Cham (2016)

\bibitem{Gaertler2005}
Gaertler, M.: Clustering. In: Network Analysis: Method. Found.

\bibitem{karypis1998fast}
Karypis, G., Kumar, V.: A fast and high quality multilevel scheme for
  partitioning irregular graphs. Scientific Computing  (1998)

\bibitem{Kastrin2014}
Kastrin, A., Rindflesch, T.C., Hristovski, D.: Link prediction on the semantic
  {MEDLINE} network - an approach to literature-based discovery. In: The
  Discovery Science Conference (2014)

\bibitem{Le2014}
Le, Q.V., Mikolov, T.: Distributed representations of sentences and documents.
  CoRR  \textbf{abs/1405.4053} (2014)

\bibitem{liben2007link}
Liben-Nowell, D., Kleinberg, J.: The link-prediction problem for social
  networks. JASIST  \textbf{58}(7) (2007)

\bibitem{newman2006modularity}
Newman, M.E.: Modularity and community structure in networks. Proceedings of
  the national academy of sciences  \textbf{103}(23) (2006)

\bibitem{Palma2014}
Palma, G., Vidal, M., Raschid, L.: Drug-target interaction prediction using
  semantic similarity and edge partitioning. In: ISWC (2014)

\bibitem{paulheim2017knowledge}
Paulheim, H.: Knowledge graph refinement: A survey of approaches and evaluation
  methods. Semantic Web Journal  \textbf{8}(3) (2017)

\bibitem{RibonVKS16}
Rib{\'{o}}n, I.T., Vidal, M., K{\"{a}}mpgen, B., Sure{-}Vetter, Y.: {GADES:}
  {A} graph-based semantic similarity measure. In: SEMANTICS (2016)

\bibitem{sachan2010using}
Sachan, M., Ichise, R.: Using semantic information to improve link prediction
  results in network datasets. IJET  \textbf{2}(4) (2010)

\bibitem{KOI2016}
Traverso-Rib{\'o}n, I., Palma, G., Flores, A., Vidal, M.E.: Considering
  semantics on the discovery of relations in knowledge graphs. In: EKAW (2016)

\bibitem{wang2015link}
Wang, P., Xu, B., Wu, Y., Zhou, X.: Link prediction in social networks: the
  state-of-the-art. Link Prediction in Social Networks(SCIS)  \textbf{58}(1)
  (2015)

\bibitem{XiaWBL17}
Xia, F., Wang, W., Bekele, T.M., Liu, H.: Big scholarly data:a survey. {IEEE}
  Big Data  (2017)

\end{thebibliography}

\end{document}